# A Back-End-Of-Line Compatible, 2-Terminals, Ferroelectric Analog Non-Volatile Memory

L. Bégon-Lours[1], M. Halter[1,2], D. Dávila Pineda[1], V. Bragaglia[1], Y. Popoff[1,2], A. La Porta[1], D. Jubin[1], J. Fompeyrine[1,3] and B. J. Offrein[1]
[1]IBM Research GmbH - Zurich Research Laboratory, CH-8803 Rüschlikon, Switzerland, email: lbe@zurich.ibm.com
[2]ETH Zürich, Zürich, Switzerland, [3]Currently at Lumiphase AG, Zürich, Switzerland

***Abstract*—** A Ferroelectric Analog Non-Volatile Memory based on a $WO_x$ electrode and ferroelectric $HfZrO_4$ layer is fabricated at a low thermal budget (~375°C), enabling BEOL processes and CMOS integration. The devices show suitable properties for integration in crossbar arrays and neural network inference: analog potentiation/depression with constant field or constant pulse width schemes, cycle to cycle and device to device variation <10%, ON/OFF ratio up to 10 and good linearity. The physical mechanisms behind the resistive switching and conduction mechanisms are discussed.

## I. Introduction

Bio-inspired analog hardware accelerators performing the synaptic function in neuromorphic computing are being developed, based on memristors: trimmable resistances from which the value can vary reversibly upon an external stimulus, in a persistent way. The choice of ferroelectric materials is attractive as the polarization can be changed by the application of an external electric field and is remnant. The growth of epitaxial thin films allowed the demonstration of ferroelectric memristors [1], however, integrating such films on CMOS remains challenging and expensive, requiring flip-chip or wafer bonding techniques. The discovery of ferroelectricity in hafnium oxide enabled the demonstration of ferroelectric memristors in the Front- and Back-End-Of-Line (BEOL) [2], [3] through techniques allowing crystallization in the ferroelectric phase with a low thermal budget [4]. Ferroelectric Tunnel Junctions (FTJs) possess only two terminals and are a desirable approach for crossbar array fabrication. An FTJ consists of a ferroelectric layer separating two different electrodes. Upon polarization reversal, the energy profile and thus the transport probability of a carrier across the junction is modified. Their fabrication is challenging as the ferroelectric layer must be thick enough to stabilize ferroelectricity in multiple configurations, but thin enough to allow electric conduction. In this work, we use an oxide semiconducting electrode, $WO_x$. We demonstrate the fabrication of an FTJ showing analog resistive switching in the BEOL with low thermal budget (~375°C) and low-cost materials (Hf, Zr, W).

## II. BEOL Fabrication

An asymmetric Metal (M) / Ferroelectric (FE) / semiconductor (SC) / Metal (M) (Fig. 1) layer stack was deposited by Atomic Layer Deposition at 300-350°C. The semiconducting layer is $WO_{3-x}$, a metal oxide with *n*-type semi-conducting properties due to the presence of oxygen vacancies. The ferroelectric layer is $HfZrO_4$ (HZO), it is capped by metallic TiN (M) to favor the crystallization of HZO in the ferroelectric phase [5] by a ms-flash lamp annealing (ms-FLA) technique (Fig. 2): the sample is heated to a moderate temperature of 375°C, then a 20 ms long flash of 70 J/cm$^2$ in energy is applied to the surface. The $I_d$-$V_g$ characteristics of P- and N-MOS (130 nm) test transistors were not affected by such treatment. W is sputtered on-top and capacitors are then defined by reactive-ion etching of the top electrode (W/TiN). Here, we demonstrate for the first time the fabrication of HZO, ferroelectric, analog non-volatile memories at a thermal budget of ~375°C. X-Ray Diffraction (Fig. 3) and X-Ray Reflectivity (Fig. 4) measurements confirm the crystallization in the orthorhombic (o-) or tetragonal (t-) phase of HZO, the absence of monoclinic phase, the polycrystalline nature of the films, the well-defined interfaces and the thickness of 4.9 nm for the dielectric barrier. Junctions do not require wake-up and Dynamic Hysteresis Measurement (DHM) on a 120 µm diameter junction shows that no breakdown is observed after $10^{10}$ switching cycles with triangular pulses of +/- 2V (Fig. 5).

## III. A Ferroelectric Memristor

In this section the resistive memory characteristics of the junction are described. Fig. 6 describes the measurement scheme: first, a pulse of amplitude $V_{write}$ and duration $t_{width}$ is applied across the junction to align the ferroelectric domains with the applied field. The polarization screening in the M and SC layers occurs over a distance inversely proportional to the carrier density in these two layers, and hence, the energy profile as well as the conductance of the junction is modified. Subsequently, an I(V) sweep in the range +/- 300 mV measures the resistance of the junction.

### *A. DC characterization*

The device characteristics are highly nonlinear with voltage (Fig. 7). The ON/OFF, defined as the ratio of the currents measured in the Low Resistive State, LRS (after applying -1.6 V) and the High Resistive State, HRS (after applying +2.4 V) is >10 for $V_{read}$=100 mV. Fig. 8 shows that the LRS, HRS and intermediate states can be reversibly reached and in a remnant way. In the LRS, the polarization points towards the SC (in accumulation mode) and a large coercive field (1.6 MV.cm$^{-1}$) is necessary to switch the ferroelectric domains. In the HRS, the ferroelectric field-effect depletes the SC layer. The poor screening in FE/SC junctions, when the SC is depleted, is usually responsible for the destabilization of the polarization in FTJs. In this work, the metal oxide electrode plays a role in the stabilization of the HRS and allows a large memory window of 1.4 V, as discussed in §IV.A.

*B. Pulse characterization*

Weight update potentiation (depression) is then demonstrated by sending sequences of negative (positive) pulses of 50 µs and increasing amplitude (Fig. 9), showing good repeatability from cycle to cycle. Cumulative Distribution Function (CDF) plots of the resistance (Fig. 10) show analog depression (LRS to HRS) and discrete potentiation (HRS to LRS, >10 levels for 25 mV steps). The memristors also demonstrate constant field weight update, by increasing the pulse duration (Fig. 11). Contrary to previous work [2] the ON/OFF is not drastically reduced compared to the increasing amplitude scheme. In Fig. 13 the normalized conductance is fitted by a $\sigma_0(1-e^{-Count/A})$ function were A is the fit parameter [6]. Interestingly linearity (see Fig. 12 and Fig. 13) is opposite for both schemes (sharp potentiation at constant $V_{ampl}$ vs sharp depression at constant $t_{width}$) showing that symmetry can be tuned using a hybrid scheme by increasing both pulse width and amplitude.

*C. Integration in a cross-bar array for inference*

On top of the low-thermal budget fabrication, the devices show a retention of >10 days (Fig.14) and are stable against heating >45°C (Fig.19). They show a small device to device variation ($\sigma$=0.1 in the HRS, Fig. 15) and scalability: current density (J) characteristics overlap for capacitors of various sizes (Fig. 16). Thanks to the high resistance, the energy of the pulse during writing is < 1 pJ. The non-linearity is high (I(V)/I(V/2)>40 for V>0.5 V) which allows built-in self-selection (limited sneak paths in absence of selectors) in a writing scheme were $V_{write}/2$ is applied to unselected rows [7]. The integration of the junctions in crossbars is limited by their high resistance: extrapolation to sub-micrometric devices leads to currents <pA. This is circumvented by reducing the HZO thickness: in a related work a $10^4$ higher conductance in the HRS is obtained with 4.5 nm thick HZO based junctions.

## IV. DISCUSSION

In order to provide guidelines for the optimization of the ferroelectric memristors for crossbar array fabrication, the physical mechanisms controlling the memory (in particular the stability of the HRS), the conductance and the non-linearity (conduction mechanisms) are explored.

*A. Role of the metal oxide electrode*

The stability of the polarization in the HRS and the dependence of the resistive switching on the pulse duration are discussed in terms of oxygen exchange between the ferroelectric and the metal oxide layer. Changing the metal oxide thickness while keeping HZO thickness constant has minor effect on the HRS and LRS, showing that the resistance is dominated by the HZO layer and indicating that the resistive switching is mainly due to ferroelectric switching. However, the dependence of the resistive switching on the pulse duration (Fig. 11) indicates that other than purely ferroelectric effects play a role in the switching mechanism. We anticipate a resistance change through field driven migration of oxygen from the SC to the FE layer, allowing oxygen vacancies in the SC layer able to screen the polarization (Fig. 18). This is supported by the observation of a strongly reduced ON/OFF (Fig. 19) when using $TiO_2$ as SC (allowing little oxygen exchange).

*B. Conduction mechanisms*

Devices are tested under temperature cycling (Fig. 20). The ON/OFF is not affected by the temperature, but the resistance decreases upon heating (Fig. 22). Such dependence and simulations discard direct tunneling as being the dominant mechanism in these junctions (Fig. 21). Log(J/V) is constant at small fields (Ohmic regime, 0~100 mV) and linear with $V^{0.5}$ at higher fields (Poole-Frenkel regime, 200~300 mV), Fig. 23. Physical parameters are extracted from the variation of the parameters of the linear regressions with the temperature. In the Poole-Frenkel regime (Fig. 24) and in the Ohmic regime (Fig. 25) a small energy barrier of 0.1~0.2 eV is measured, pointing toward conduction via oxygen vacancies in the HZO layer. The oxygen content in HZO is controlled by the growth and annealing conditions and is an additional knob (with the thickness) to tune the conductance of the junctions.

## V. CONCLUSION

The ferroelectric analog non-volatile memory technology presented in this works shows characteristics comparable to other technologies (Fig. 26). The low-thermal budget process, makes it a promising candidate for the fabrication of crossbar arrays for deep-neural networks accelerators. The devices show analog potentiation/depression with constant field or constant pulse width schemes, which makes them especially interesting for inference where such schemes are easily implemented. This first generation of BEOL, ferroelectric 2-terminals memristors shows non-linearity of (1.9/-4), ON/OFF ratio up to 10 and cycle to cycle and device to device variation <10%. The future optimization of the devices (with thinner HZO layer) is led by the understanding of the memory and conduction mechanisms, with focus on oxygen exchange between the electrode and the ferroelectric layer as well as conduction mediated by oxygen vacancies though the dielectric.

ACKNOWLEDGMENT

This work is funded by H2020 "FREEMIND" (840903) and is based on the work presented at IEEE EDTM 2021.

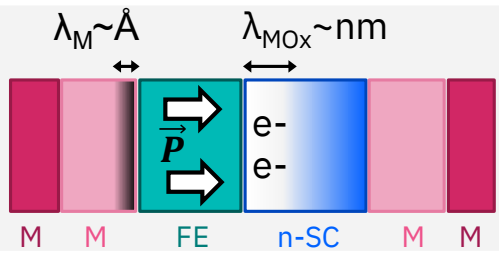

Fig. 1. FTJ with a metallic (M) and n-type metal oxide semiconducting (n-SC) electrodes. SC has a small carrier density and a large screening length.

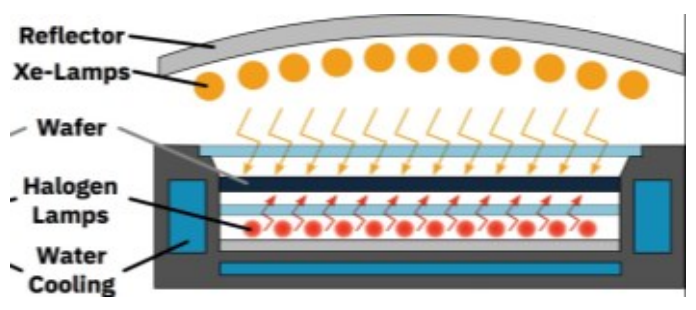

Fig. 2. Crystallization of HZO at only 375°C using a ms-flash lamp annealing technique. Critical step ensuring CMOS compatibility in BEOL.

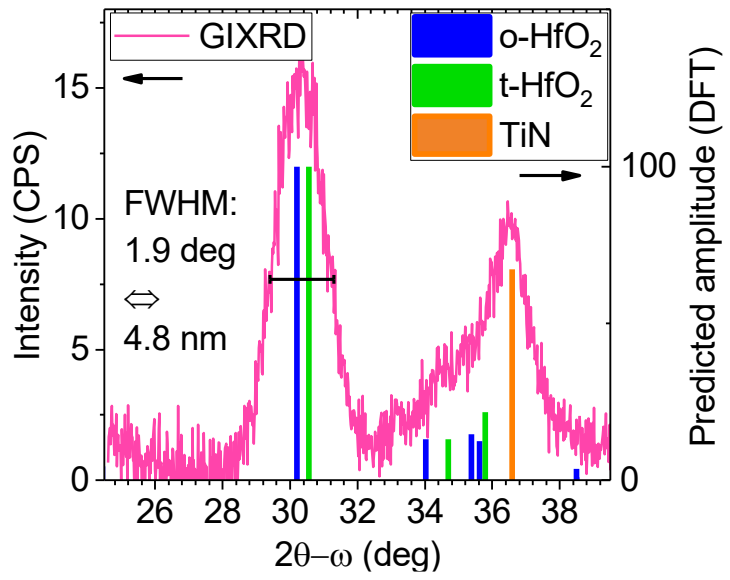

Fig. 3. GIXRD: HZO crystallization in a non-monoclinic phase.

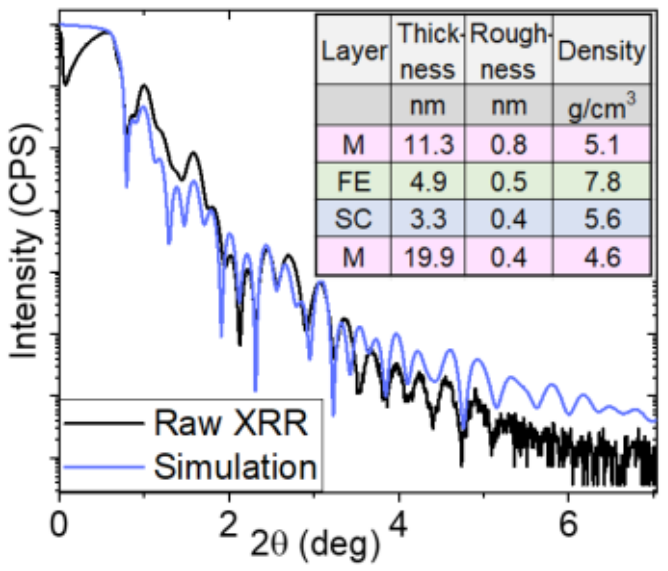

Fig. 4. X-Rays Reflectivity: sharp interfaces and HZO thickness of 4.9 nm.

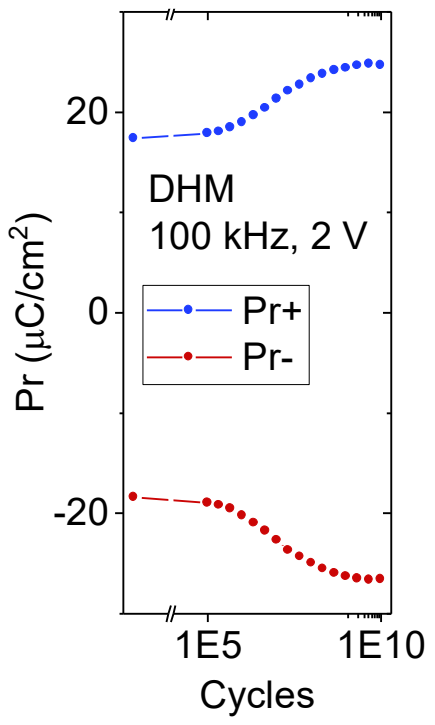

Fig. 5 HZO polarization can be switched for more than $10^{10}$ cycles. No wake-up is needed.

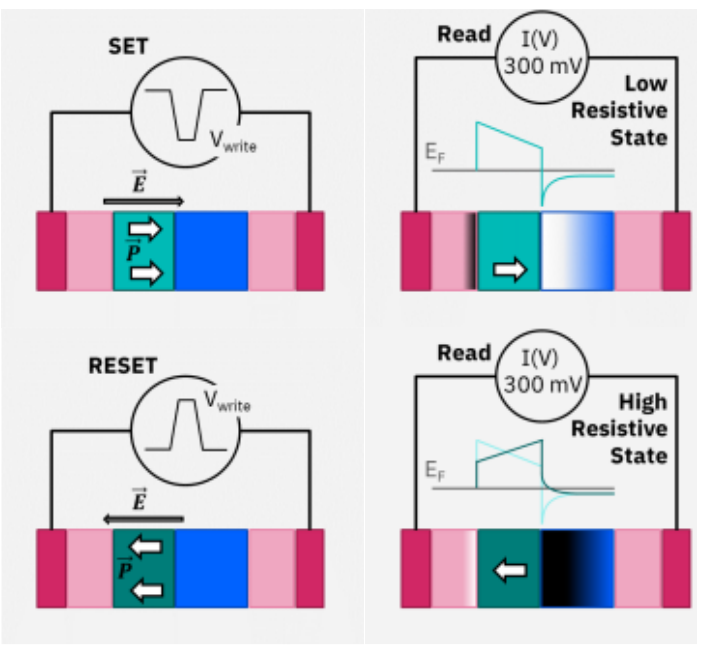

Fig. 6. Measurement scheme of the resistive switching. Write pulses change the polarization of the FE layer (n-SC layer is grounded). The change in the energy profile is measured by an I(V) sweep.

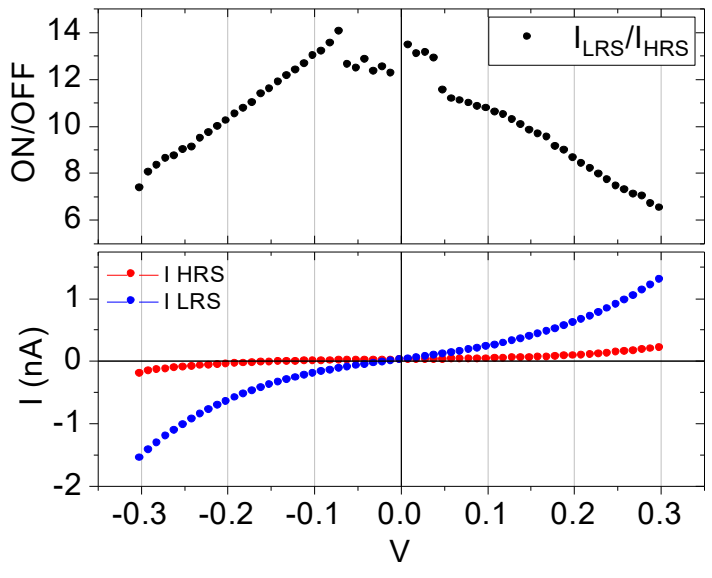

Fig. 7. 120 µm Ø junction: I(V) are non-linear. ON/OFF=I(LRS)/I(HRS) reaches 10 at a read voltage of 100 mV. n-SC layer is grounded.

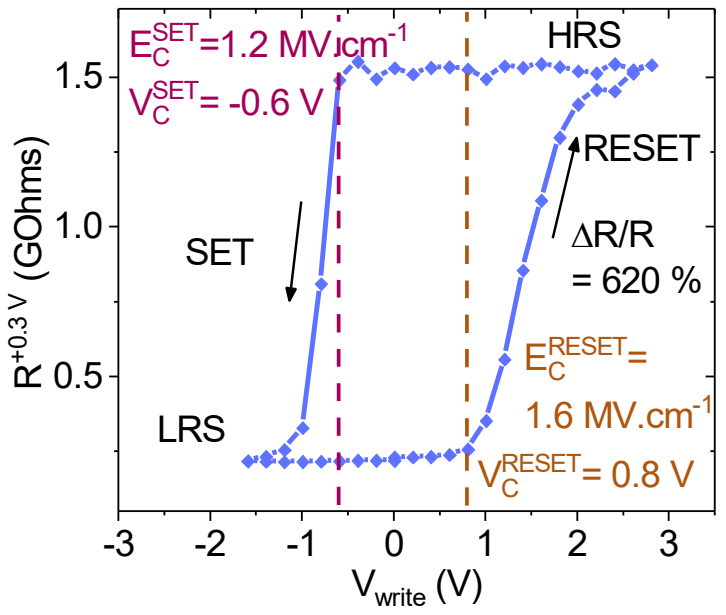

Fig. 8. Read resistance $R^{+0.3V}$ after DC writing $V_{write}$. Very stable HRS ($V_C^{SET}$= -0.6V) despite screening with depleted SC.

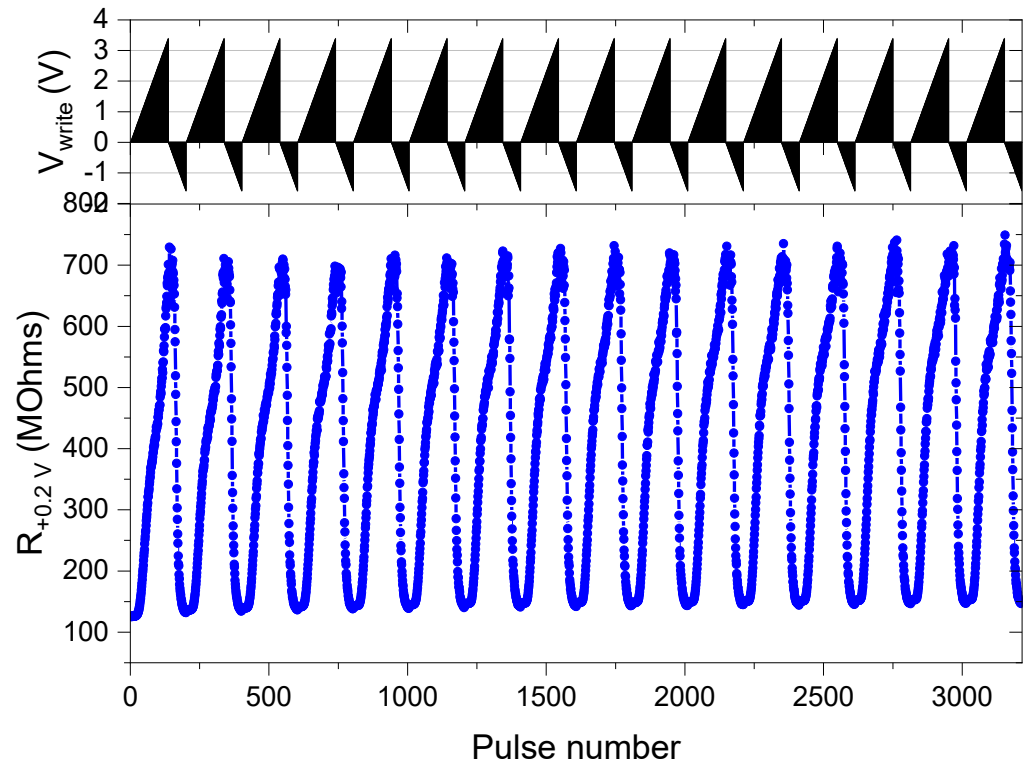

Fig. 9. Potentiation/Depression over 17 cycles with pulses of constant duration (50 us) and increasing amplitude.

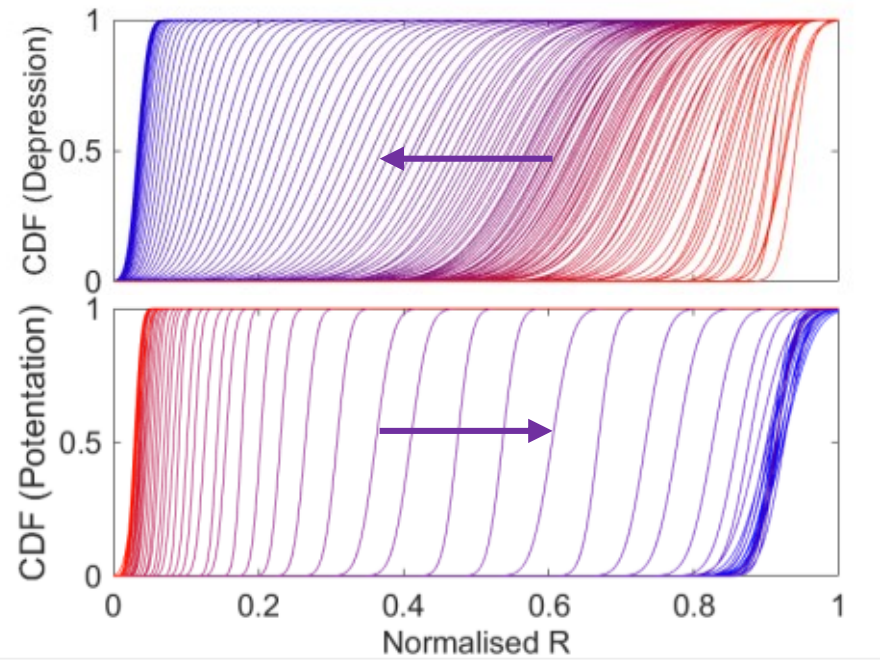

Fig. 10. CDF plots of the resistance (increasing amplitude scheme): analog depression and discrete potentiation.

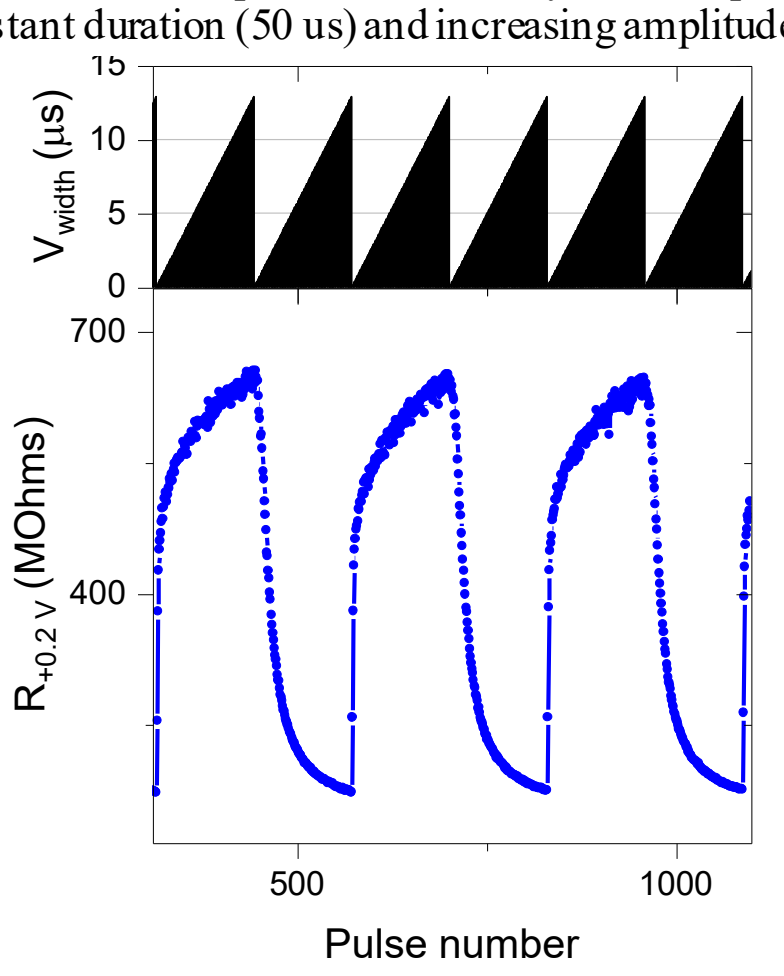

Fig. 11. Potentiation/Depression: pulses of constant field $V_{write}$=3.2 V (-1.4V) and increasing duration $V_{width}$.

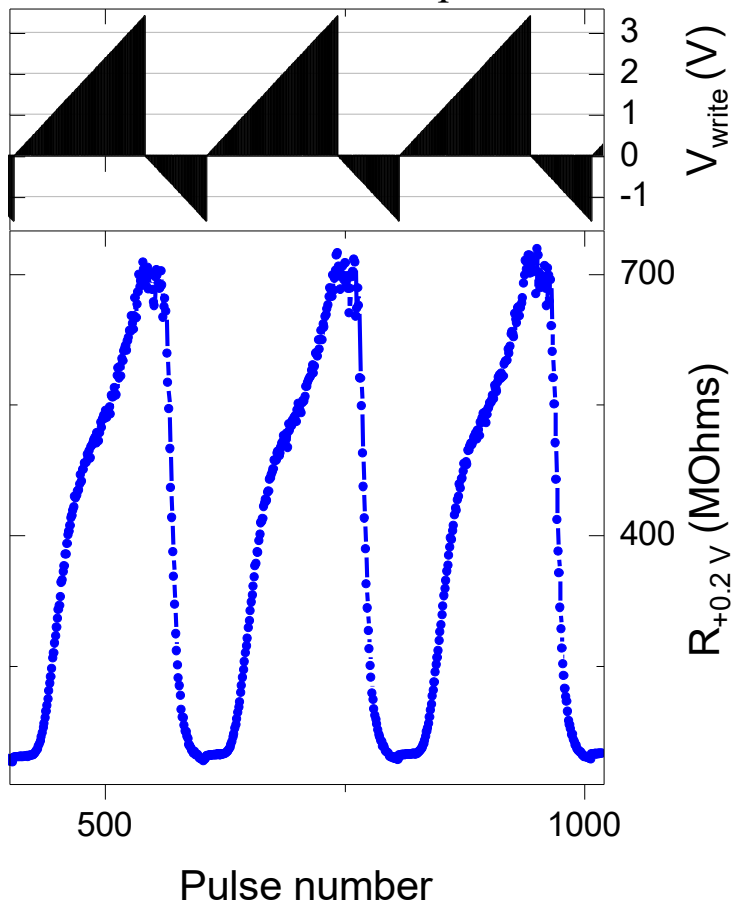

Fig. 12. Potentiation/Depression: pulses of constant duration (50 us) and increasing amplitude, zoom in.

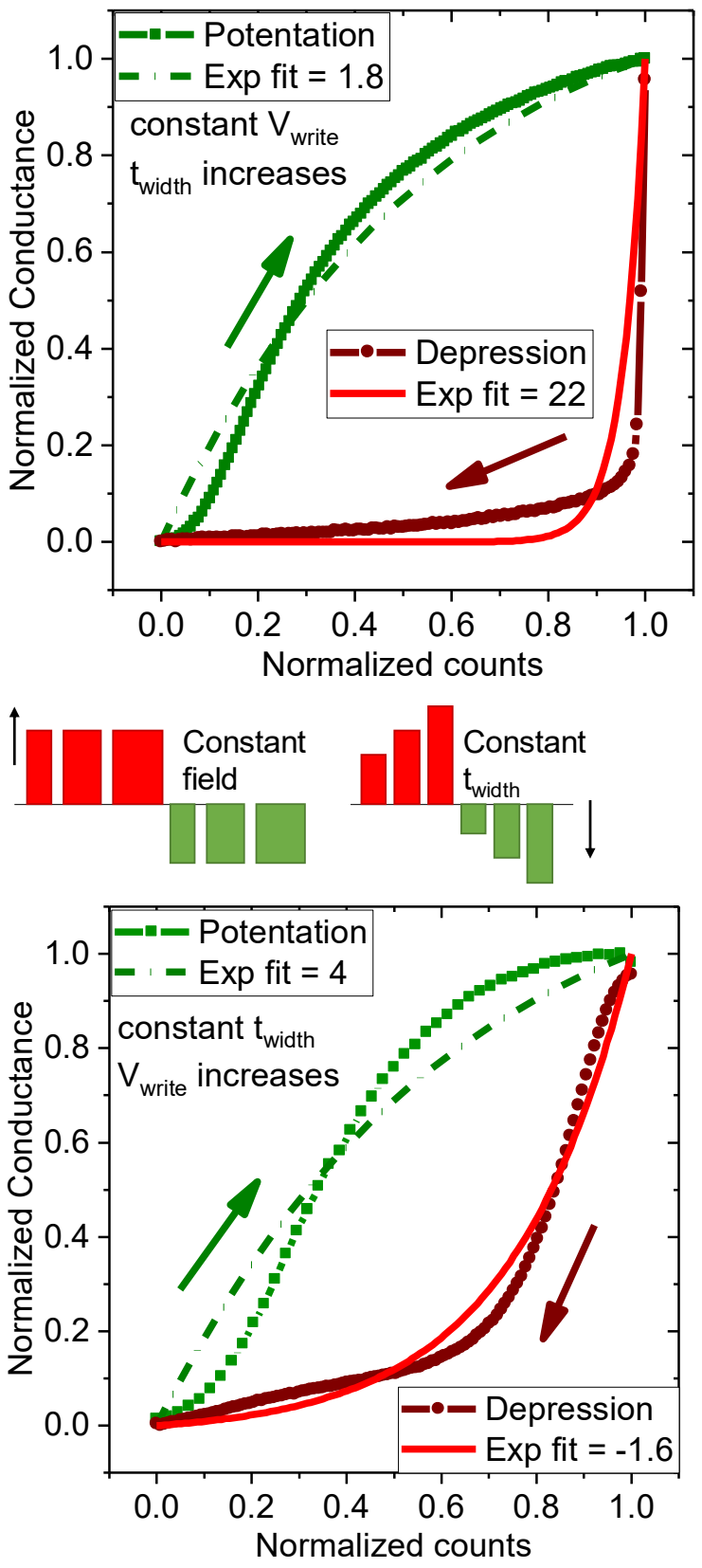

Fig. 13. Normalized conductance vs counts (pulse number). Exp. fit parameters is measured as in ref. [6]. Up: for constant field, down: for constant pulse duration.

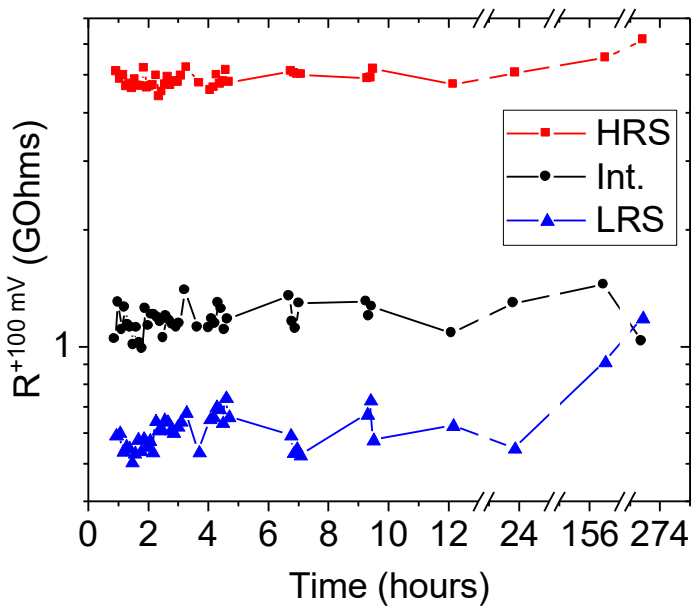

Fig. 14. Junctions measured after set in HRS, LRS and intermediate state up to 11 days.

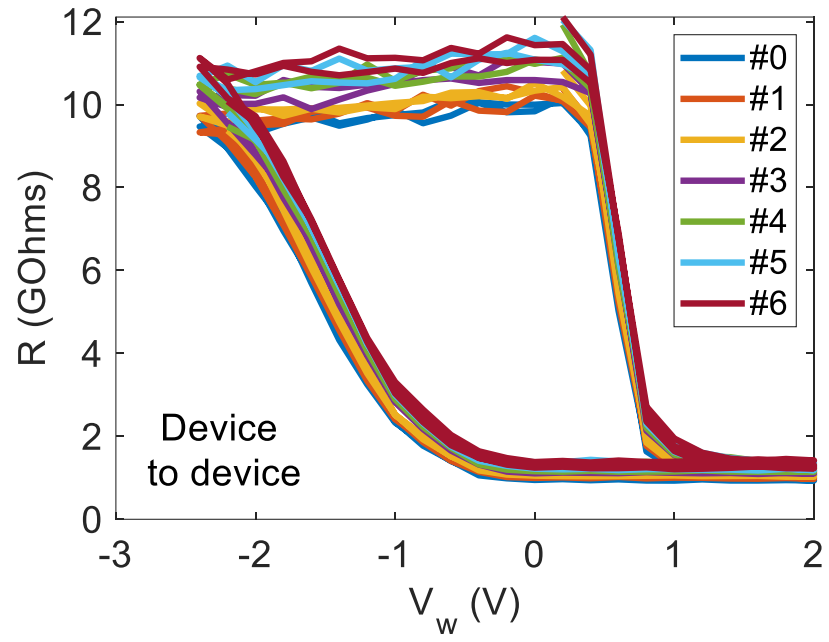

Fig. 15. Small device to device variation. n-SC layer is biased.

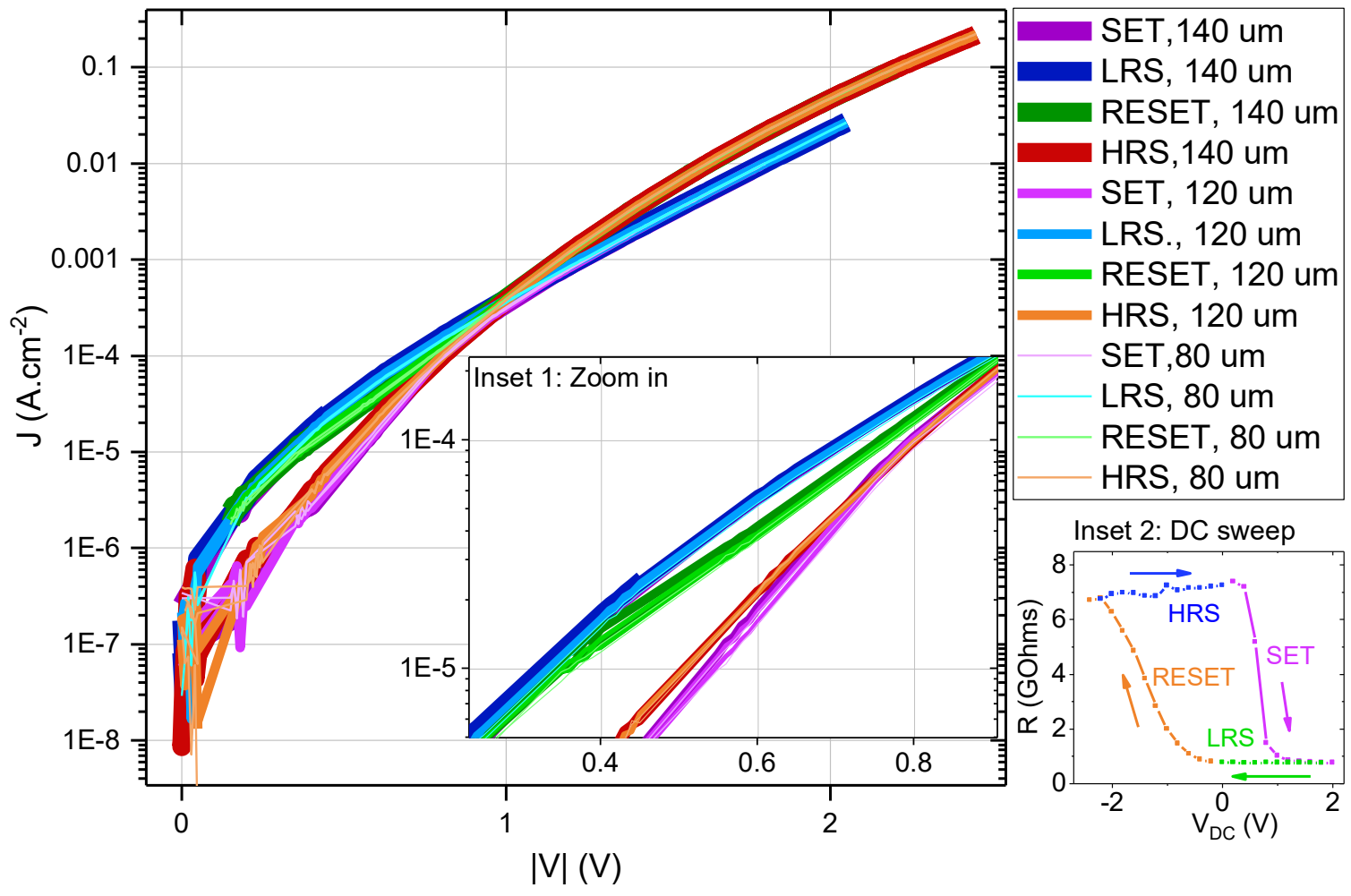

Fig. 16. J(V) characteristics identical for different sizes: homogeneous conduction (no filament). n-SC layer is biased.

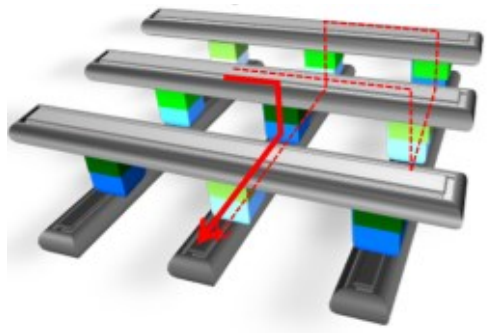

Fig. 17. Sneak paths in crossbars are limited by I(V) non-linearity

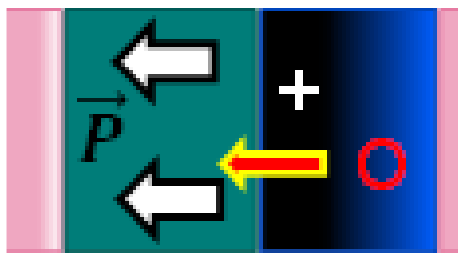

Fig. 18. Expected mechanism for HRS stabilization.

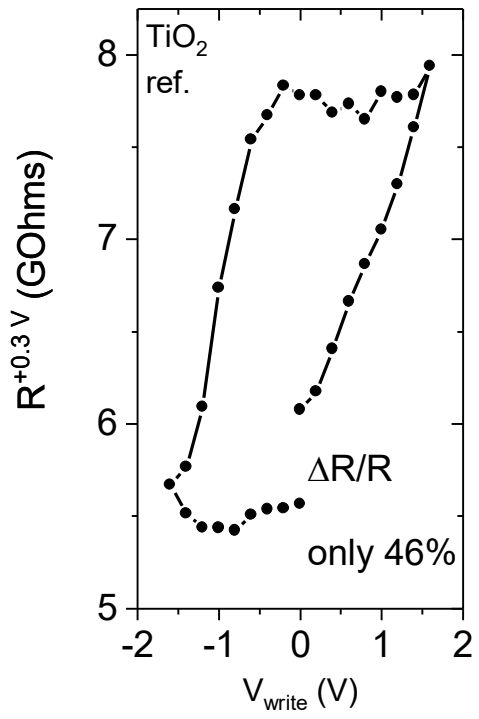

Fig. 19. Poor ON/OFF of TiO based junctions pointing to the role of the MOx electrode.

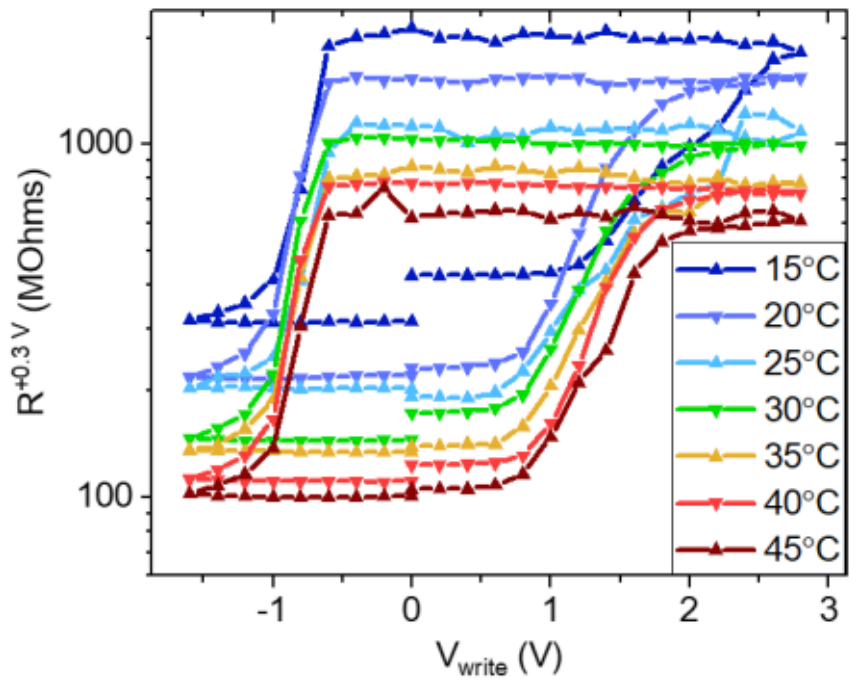

Fig. 20. R($V_{write}$) loops at various temperatures. n-SC layer is grounded.

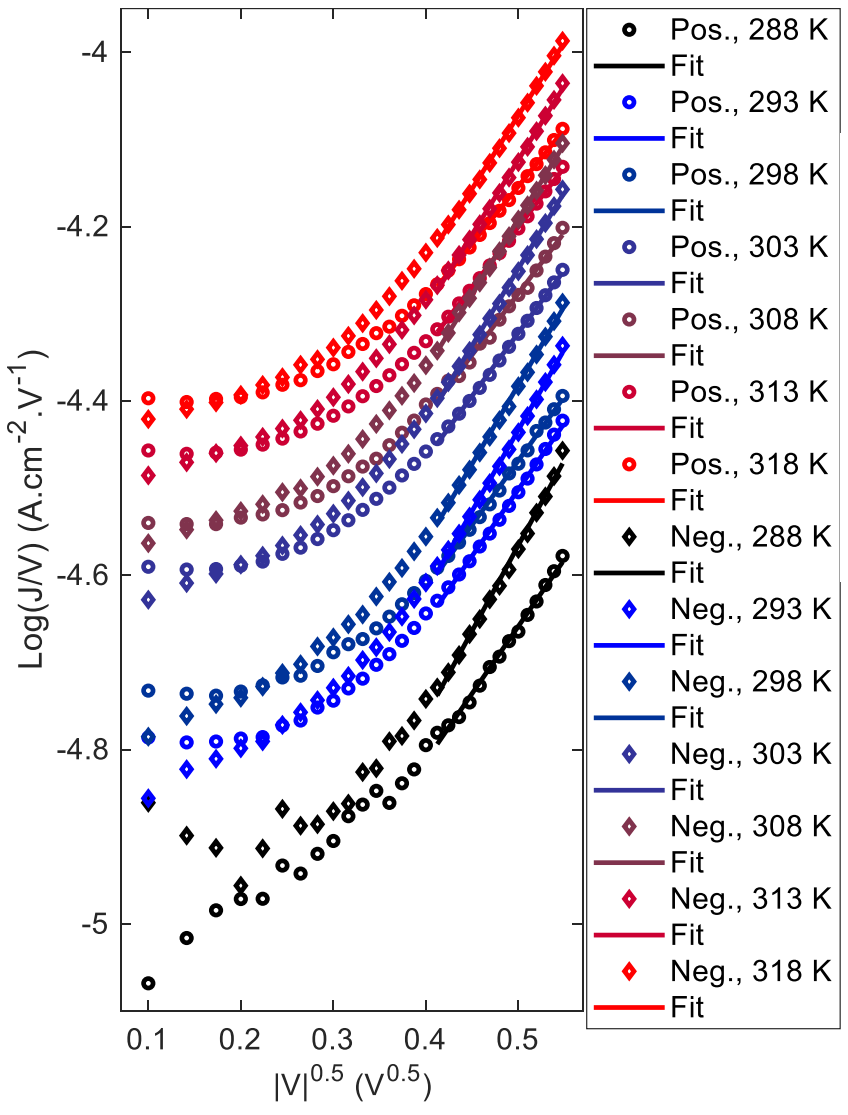

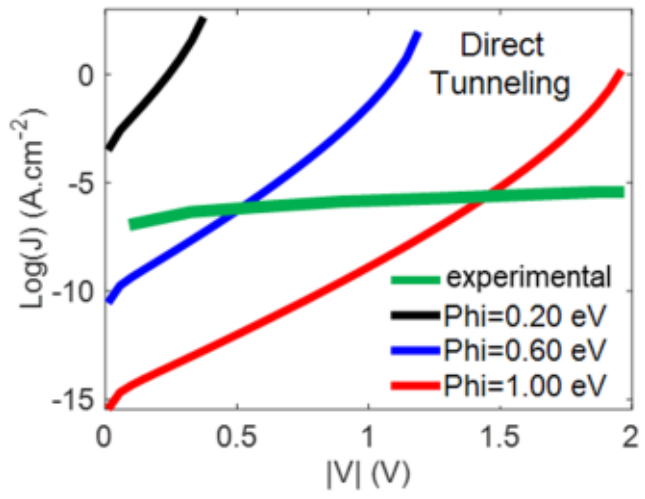

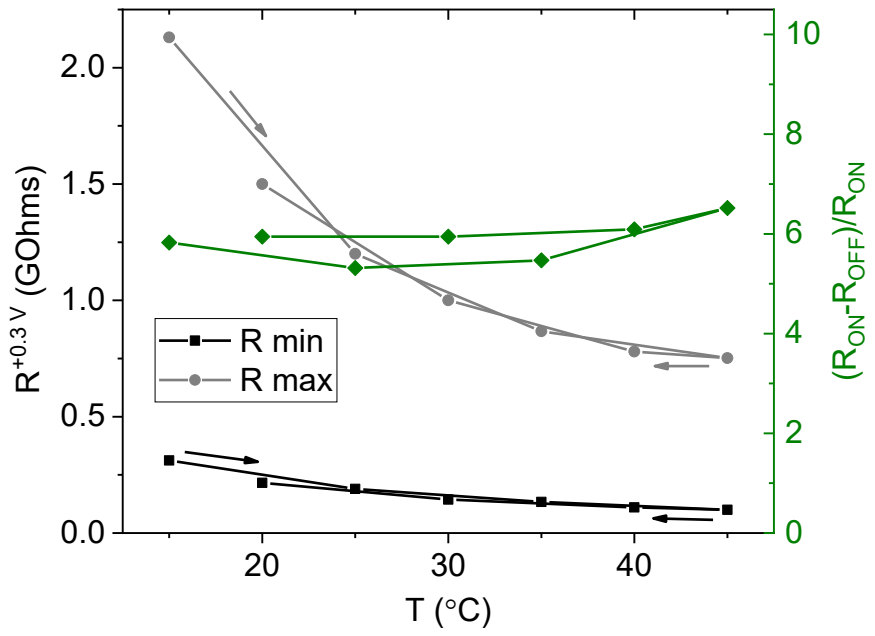

Fig. 23. Log(J/V)=f($V^{0.5}$) at different temperatures and linear fits in the 200~300 mV range.

Fig. 21. Experiment vs simulation demonstrates direct tunneling is not the main mechanism.

Fig. 22. Heating modifies reversibly the resistance but ON/OFF stays constant.

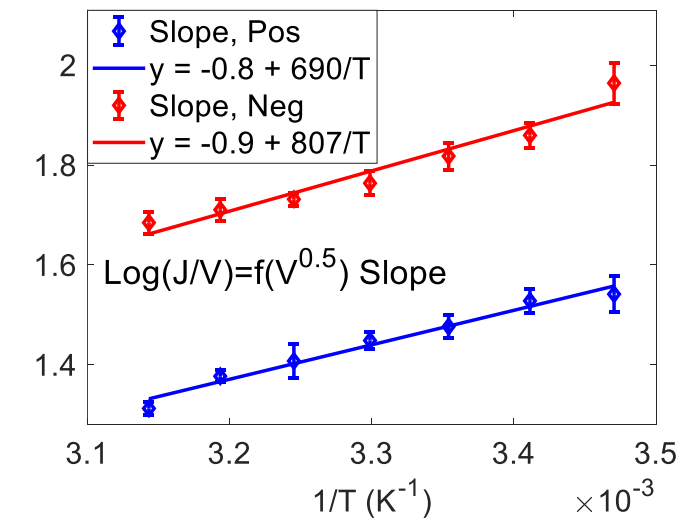

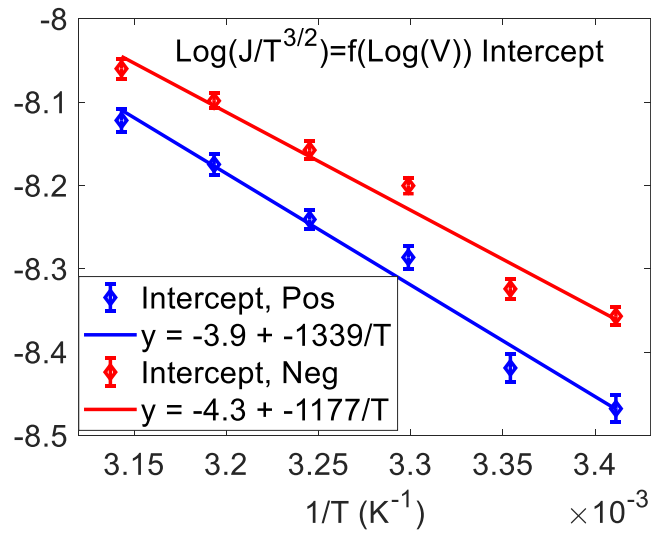

| Device type | [8] MO-ECRAM | TaOx/HfOx [9] | PCMO [10] | AlOx/ HfO2 [11] | c-FSJ [12] | This work |
|---|---|---|---|---|---|---|
| Nonlinearity | <1 | 0.04/-0.63 | 3.7/-6.8 | 1.94/-0.61 | 4.2/-4.2 | 1.9/-4.3 |
| RON | 67KΩ | 100KΩ | 23MΩ | 16.9KΩ | 100KΩ | 100MΩ |
| ON/OFF | 20 | 10 | 6.84 | 4.43 | 21 | 7 |
| Depression | 4V/ 10ns | 1.6V/50ns | 2V/1ms | 0.9V/ 100µs | 2V/80ns | 2.4V/ 50 us |
| Potentation | -4V/ 10ns | -1.6V/ 50ns | -2V/ 1ms | -1V/ 100µs | -2V/ 80ns | -1.6V/ 50us |
| Cycle-to-cycle var. | <10% | 3.70% | <1% | 5% | <0.5% | 10% |
| Area (µm$^2$) | 40 | 8663.1 | 6292.3 | 21,846 | 184,420 | 14,400 |

Fig. 24. Temperature dependence of the slope of the Log(J/V)=f($V^{0.5}$) linear regression (Poole-Frenkel regime).

Fig. 25. Temperature dependence of the intercept of the Log(J/$T^{3/2}$)=f(Log(V) linear regression (Ohmic regime).

Fig. 26. Benchmark.